\documentclass[epj]{webofc}
\usepackage[varg]{txfonts}   
\wocname{EPJ Web of Conferences}
\woctitle{ICNFP 2017}
\usepackage{color} 
\usepackage{amsfonts}
\usepackage{amsbsy}
\usepackage{mathrsfs}
\usepackage{graphicx}
\def\lsim{\mathrel{\rlap{
\lower4pt\hbox{\hskip-3pt$\sim$}}
    \raise1pt\hbox{$<$}}}     
\def\gsim{\mathrel{\rlap{
\lower4pt\hbox{\hskip-3pt$\sim$}}
    \raise1pt\hbox{$>$}}}     
\def\scr#1{\mbox{\scriptsize #1}}

\begin{document}
\selectlanguage{english}
\title{Three-fluid Hydrodynamics-based Event Simulator Extended by UrQMD final State interactions (THESEUS)
for FAIR-NICA-SPS-BES/RHIC energies}%
\author{P. Batyuk\inst{1}
        \and
        D. Blaschke\inst{2,3,4}
        \and
        M. Bleicher\inst{5,6,7}
        \and
        Yu.B. Ivanov\inst{3,4,8}\fnsep\thanks{speaker, e-mail: Y.Ivanov@gsi.de} \and
        Iu. Karpenko\inst{9,10}
        \and
        L. Malinina\inst{1,11}
        \and
        S. Merts\inst{1}
        \and
        M. Nahrgang\inst{12}
        \and
        H. Petersen\inst{5,6,7}
        \and
        O. Rogachevsky\inst{1}
        }

\institute{Veksler and Baldin Laboratory of High Energy Physics, JINR Dubna, 141980 Dubna, Russia
\and
Institute of Theoretical Physics, University of Wroclaw, 50-204 Wroclaw, Poland
\and
Bogoliubov Laboratory of Theoretical Physics, JINR Dubna, 141980 Dubna, Russia
\and
National Research Nuclear University "MEPhI", 
115409 Moscow, Russia 
\and
Frankfurt Institute for Advanced Studies (FIAS),
Frankfurt am Main, Germany
\and
Institut f\"ur Theoretische Physik, Goethe Universit\"at, 
Frankfurt am Main, Germany
\and
GSI Helmholtzzentrum f\"ur Schwerionenforschung GmbH, 
Darmstadt, Germany
\and
National Research Centre "Kurchatov Institute" 
123182 Moscow, Russia
\and
Bogolyubov Institute for Theoretical Physics, 03680 Kiev, Ukraine
\and
INFN - Sezione di Firenze, I-50019 Sesto Fiorentino (Firenze), Italy
\and
Skobeltsyn Institute for Nuclear Physics, Lomonosov Moscow State University, Moscow, Russia
\and
SUBATECH, UMR 6457 (IMT-Atlantique, Universit\'{e} de Nantes, CNRS-IN2P3), 44307 Nantes, France
Nantes cedex 3, France
}

\abstract{
We present a new event generator based on the three-fluid hydrodynamics (3FH) approach, 
followed by a particlization at the hydrodynamic decoupling surface and a subsequent UrQMD afterburner stage
based on the  microscopic UrQMD transport model that accounts for hadronic final state interactions.
First results for  Au+Au collisions are presented.
The following topics are addressed: the directed flow, transverse-mass spectra, as well as rapidity distributions 
of protons,  pions and kaons for two model equations of state, 
one with a first-order phase transition, the other with a crossover transition. 
Preliminary results on the femtoscopy are also discussed.%
We analyze the accuracy of reproduction of the 3FH results by the new event generator
and the effect of the subsequent UrQMD afterburner stage. 
}
%
%
\maketitle

\section{Introduction}

The onset of deconfinement in relativistic heavy-ion collisions 
and the search for a critical endpoint 
is now in the focus of 
theoretical and experimental studies of the equation of state (EoS) and the phase diagram 
of strongly interacting matter. 
This challenge is one of the main motivations for 
the currently running beam-energy scan (BES) at the Relativistic Heavy-Ion Collider (RHIC)
at Brookhaven National Laboratory (BNL) \cite{Stephans:2006tg} 
and at the CERN Super-Proton-Synchrotron (SPS) \cite{SPS-scan} as well as
for constructing the
Facility for Antiproton and Ion Research (FAIR) in Darmstadt \cite{FAIR} and the
Nuclotron-based Ion Collider fAcility (NICA) in Dubna \cite{NICA}. 

Three-fluid hydrodynamics (3FH) \cite{Ivanov:2005yw} was designed to simulate heavy-ion collisions at moderately relativistic energies, i.e. precisely in the energy range of the 
expected onset of deconfinement. 
In recent years applications of 3FH demonstrated a strong preference of deconfinement scenarios for the explanation of available experimental data 
\cite{Ivanov:2013wha,Ivanov:2012bh,Ivanov:2013yqa,Ivanov:2013yla,Konchakovski:2014gda,Ivanov:2014zqa,Ivanov:2014ioa,Ivanov:2016sqy}. 
However, up to now 3FH has been facing certain problems. 
From the theoretical side, the model lacks an afterburner stage that can 
play an important role for some observables. 
From the practical point of view, the model was not well suited for data simulations in terms of experimental events, because the model output consisted of fluid characteristics rather than of 
a set of observable particles.  

In this contribution, we present a new Three-fluid Hydrodynamics-based Event Simulator Extended 
by UrQMD final State interactions (THESEUS) \cite{Batyuk:2016qmb}
and its application to the description of heavy-ion collisions in the FAIR-NICA-SPS-BES/RHIC energy range.
This simulator provides a solution to both the above-mentioned problems.
It presents the 3FH output in terms of a set of observed particles 
and the afterburner can be run starting from this output by means of the UrQMD model 
\cite{Bass:1993em,Bass:1998ca}. 
Thus THESEUS as a new tool allows to discuss the multifaceted physics challenges at FAIR-NICA-SPS-BES/RHIC energies.
The new simulation program has the unique feature that it can describe a hadron-to-quark matter transition of the first order which proceeds in the baryon stopping regime that is not accessible to previous simulation programs designed for higher energies.
Besides this, with THESEUS one can address
practical questions like the influence of hadronic final state interactions and of the detector acceptance, which are necessary to  focus on important physics questions. These deal with a potential discovery and investigation of the first-order phase transition line, where during a heavy-ion collision the EoS reaches its softest point \cite{Hung:1997du}. It remains an open question how this characteristic feature of the EoS manifests itself
in observables such as flow, proton rapidity distributions and femtoscopic radii. 
Particular emphasis is on the robustness of the "wiggle" \cite{Ivanov:2015vna} in the energy scan of the midrapidity curvature of the proton rapidity distribution that has been suggested as a possible signal for a first order phase transition, expected just in the range of energies at NICA and FAIR experiments. 

At present THESEUS is not an integrated approach. 
The simulation proceeds in two steps: first, a numerical solution of the 3-fluid hydrodynamics is computed with the corresponding code. 
Based on the output of the hydrodynamic part, a Monte Carlo procedure is used to sample the ensemble of hadron distributions and the UrQMD code is engaged to calculate final state hadronic rescatterings, as will be explained below.
Another present limitation which we leave for future work is the absence of event-by-event hydrodynamic evolution. 
Therefore later by an event we mean a Monte Carlo sampled set of final hadrons, which correspond to the same (average) hydrodynamic evolution.

\section{Description of the event generator THESEUS}
\label{THESEUS}
\subsection{The 3FH model} 
\label{3FD}

The 3FH model treats the collision process from the 
very beginning, i.e. from the stage of cold nuclei, up to the particlization   
from the fluid dynamics. 
This model is a straightforward extension of the two-fluid model with radiation
of direct pions \cite{Mishustin:1988mj,Mishustin:1991sp}
and of the (2+1)-fluid model of 
Refs.~\cite{Katscher:1993xs,Brachmann:1997bq}.
The 3-fluid approximation is a minimal way to simulate the finite
stopping power at the initial stage of the collision. 
Within the 3-fluid approximation a generally nonequilibrium distribution of baryon-rich
matter is simulated by counter-streaming baryon-rich fluids
initially associated with constituent nucleons of the projectile (p)
and target (t) nuclei. 
Therefore, the initial conditions for the fluid evolution are two Lorentz contracted spheres with radii of corresponding nuclei and zero diffuseness, baryon density $n_B = 0.15$ fm$^{-3}$ and energy density $m_N n_B \simeq 0.14$ GeV/fm$^3$.
In addition, newly produced particles, populating
the mid-rapidity region, are associated with a fireball (f) fluid. 
Each of these fluids is governed by conventional hydrodynamic
equations.  
The continuity equations for the baryon charge read 
   \begin{eqnarray}
   \label{eq8}
   \partial_{\mu} J_{\alpha}^{\mu} (x) &=& 0,
   \end{eqnarray}
for $\alpha=$p and t, where
$J_{\alpha}^{\mu}=n_{\alpha}u_{\alpha}^{\mu}$ is the baryon
current defined in terms of proper (i.e. in the local rest frame) net-baryon density $n_{\alpha}$ and
 hydrodynamic 4-velocity $u_{\alpha}^{\mu}$ normalized as
$u_{\alpha\mu}u_{\alpha}^{\mu}=1$. Eq.~(\ref{eq8}) implies that
there is no baryon-charge exchange between p-, t- and f-fluids, as
well as that the baryon current of the fireball fluid is
identically zero, $J_{\scr f}^{\mu}=0$, 
by construction. 
Equations of the energy--momentum exchange between fluids are formulated
in terms of 
energy--momentum tensors $T^{\mu\nu}_\alpha$ of the 
fluids
   \begin{eqnarray}
   \partial_{\mu} T^{\mu\nu}_{\scr p} (x) &=&
-F_{\scr p}^\nu (x) + F_{\scr{fp}}^\nu (x),
   \label{eq8p}
\\
   \partial_{\mu} T^{\mu\nu}_{\scr t} (x) &=&
-F_{\scr t}^\nu (x) + F_{\scr{ft}}^\nu (x),
   \label{eq8t}
\\
   \partial_{\mu} T^{\mu\nu}_{\scr f} (x) &=&
- F_{\scr{fp}}^\nu (x) - F_{\scr{ft}}^\nu (x)
+
\int d^4 x' \delta^4 \left(\vphantom{I^I_I} x - x' - U_F
(x')\tau_f\right)
 \left[F_{\scr p}^\nu (x') + F_{\scr t}^\nu (x')\right],
   \label{eq8f}
   \end{eqnarray}
where the $F^\nu_\alpha$ are friction forces originating from
inter-fluid interactions. $F_{\scr p}^\nu$ and $F_{\scr t}^\nu$ in
Eqs.~(\ref{eq8p})--(\ref{eq8t}) describe energy--momentum loss of the 
baryon-rich fluids due to their mutual friction. A part of this
loss $|F_{\scr p}^\nu - F_{\scr t}^\nu|$ is transformed into
thermal excitation of these fluids, while another part $(F_{\scr
p}^\nu + F_{\scr t}^\nu)$ gives rise to particle production into
the fireball fluid (see Eq.~(\ref{eq8f})). $F_{\scr{fp}}^\nu$ and
$F_{\scr{ft}}^\nu$ are associated with friction of the fireball
fluid with the p- and t-fluids, respectively. 
Here $\tau_f$ is the formation time, and
   \begin{eqnarray}
   \label{eq14}
U^\nu_F (x')=
\frac{u_{\scr p}^{\nu}(x')+u_{\scr t}^{\nu}(x')}%
{|u_{\scr p}(x')+u_{\scr t}(x')|}
   \end{eqnarray}
is the  4-velocity of the free propagation of the produced fireball 
matter.
In fact, this is the velocity of the fireball matter at the moment of its production. 
According to Eq.~(\ref{eq8f}),  this matter gets formed only 
after the time span $U_F^0\tau_f$ upon the production, and in a 
different space point ${\bf x}' - {\bf U}_F (x') \ \tau_f$, as
compared to the production point ${\bf x}'$. 
 The friction between fluids
was fitted to reproduce the stopping power observed in proton
rapidity distributions for each EoS, as it is described in Refs. 
\cite{Ivanov:2005yw,Ivanov:2013wha} in detail.

\begin{figure}[!hbt]
\centering
\sidecaption
\includegraphics[width=0.32\textwidth]{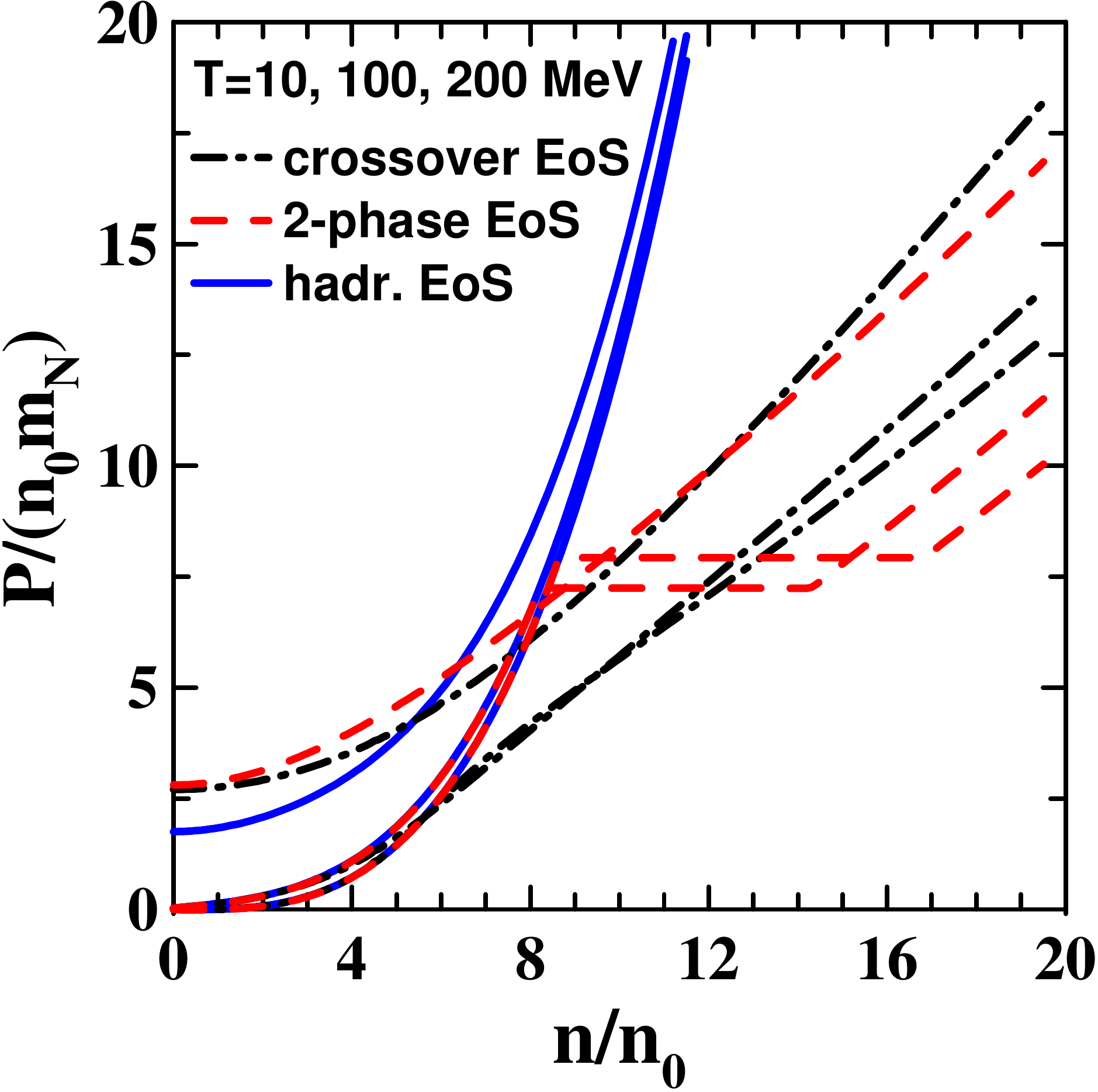}
 \caption{{
Pressure scaled by the product of normal nuclear density ($n_0=$ 0.15 fm$^{-3}$) and 
nucleon mass ($m_N$) versus baryon density scaled by the normal nuclear density
for three considered equations of state. Results are presented for three different
temperatures $T=$ 10, 100 and 200 MeV (from bottom upwards for corresponding curves).  
}} 
\label{fig1}
\end{figure}

Different equations of state (EoS) can be applied within the 3FH model. 
The recent series of simulations 
\cite{Ivanov:2013wha,Ivanov:2012bh,Ivanov:2013yqa,Ivanov:2013yla,Konchakovski:2014gda,Ivanov:2014zqa,Ivanov:2014ioa,Ivanov:2016sqy}
was performed  
employing three different types of EoS: a purely hadronic EoS   
\cite{gasEOS} (hadr. EoS) and two versions of the EoS involving    
deconfinement  \cite{Toneev06}. 
The latter two versions are an EoS with a first-order phase transition (2-phase EoS) 
and one with a smooth crossover transition (crossover EoS). 
Figure \ref{fig1} illustrates the differences between the three considered EoS. 
%

\begin{figure*}[!ht]
\centering
\includegraphics[width=14.0cm]{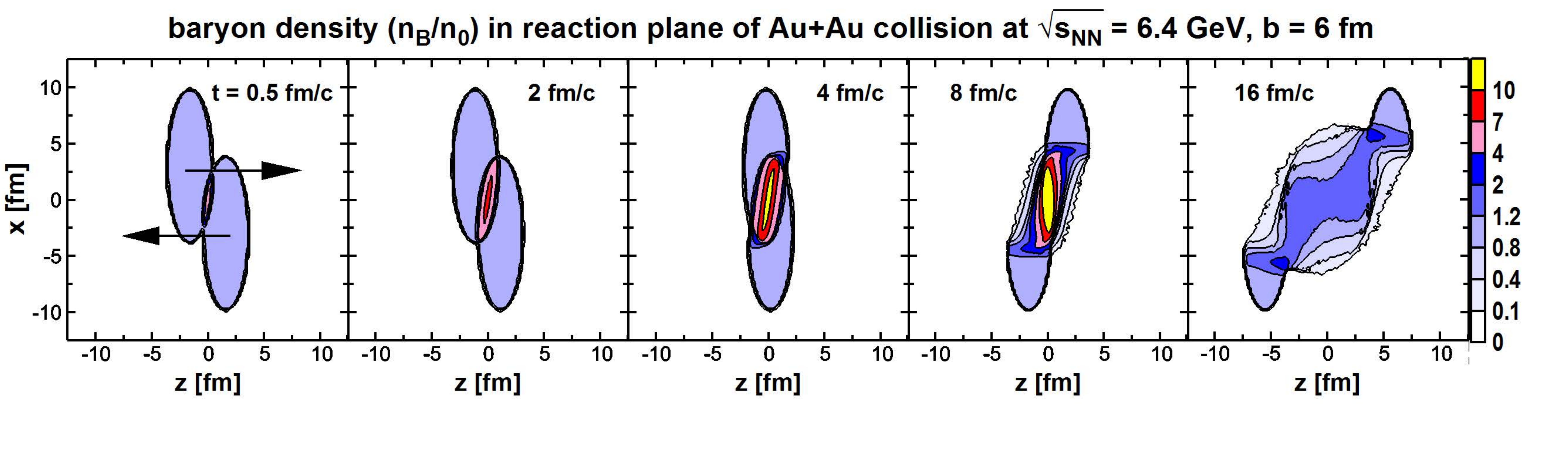}
 \caption{
Evolution of the proper  
baryon density ($n_B/n_0$) scaled by the the normal nuclear density ($n_0=0.15~$fm$^{-3}$)
in the reaction plane for a semi-central ($b=6~$fm) Au+Au collision at $\sqrt{s_{NN}}=6.4~$GeV.
} 
\label{fig2}
\end{figure*}

An application of the 3FH model is illustrated in Fig.~\ref{fig2} 
where the evolution of the proper (i.e. in the local rest frame) 
baryon density in the reaction plane is presented for a semi-central 
(impact parameter $b=$ 6 fm) Au+Au collision at $\sqrt{s_{NN}}=$ 6.4 GeV
($E_{\rm lab}= 20 $ A GeV).  
The simulation was performed with the crossover 
EoS without freeze-out. 
As can be seen from that figure, very high baryon densities are reached 
in the central region of the colliding system.

The freeze-out criterion used in the 3FH model is $\varepsilon < \varepsilon_{\scr{frz}}$, 
where $\varepsilon$ is the total energy density of all three fluids in their common rest 
frame.
More details can be found in Refs.~\cite{71,74}
The freeze-out energy density $\varepsilon_{\scr{frz}}=0.4~$GeV/fm$^3$ 
was chosen mostly on the condition of the best reproduction 
of secondary particle yields for all considered scenarios, see \cite{Ivanov:2005yw}.  
An important feature of the 3FH freeze-out is an antibubble prescription, preventing the formation of
bubbles of frozen-out matter inside the dense matter while it is still hydrodynamically evolving. 
The matter is allowed to be frozen out only if 
it is located near the border with the vacuum (this piece of matter gets locally frozen out). 
The thermodynamic quantities of the frozen-out
matter are recalculated from the in-matter EoS, with which
the hydrodynamic calculation runs, to the hadronic gas EoS%
\footnote{
In this gas EoS 48 different hadronic
species are taken into account. Each hadronic species includes all the
relevant isospin states; e.g., the nucleon species includes protons and
neutrons.}.
This is done because a part of the energy is still accumulated in
collective mean fields at the freeze-out instant. 
This mean-field energy needs to be released before entering the hadronic cascade 
in order to facilitate energy conservation. 

The output of the model is recorded in terms of Lagrangian test particles (i.e. fluid droplets)
for each fluid $\alpha$ (= p, t or f).
Each particle contains information on space-time coordinates ($t,\bf{x}$) of the frozen-out matter, 
proper volume of the test particle of the $\alpha$ fluid ($V_\alpha^{\rm pr}$), 
hydrodynamic velocity ($u^{\mu}_\alpha$) in the frame of computation, 
temperature ($T_{\alpha}$), baryonic ($\mu_{\rm B\alpha}$)  and strange ($\mu_{\rm S\alpha}$) chemical potentials.

\subsection{Particlization}
\label{Particlization}

In the multi-fluid approach one simulates the heavy ion collision from its very first moment using fluid dynamics. 
However, once the system becomes too dilute, the fluid approximation loses its applicability and individual particles are the relevant degrees of freedom. 
The process of changing from a fluid to a particle description is called "particlization" \cite{Huovinen:2012is}. 
Since we supplement the 3FH with a hadronic cascade, the particlization is not freeze-out anymore.
By definition there are only resonance decays after freeze-out, whereas in the present generator final state hadronic rescattering processes are simulated as well using the UrQMD code.

The particlization criterion is chosen to be the same as freeze-out criterion in \cite{Ivanov:2005yw}, e.g.,
$$\varepsilon_{\rm tot}<\varepsilon_{\rm frz},$$
where $\varepsilon_{\rm tot}$ is defined as:
$$\varepsilon_{\rm tot}=T^{*00}_{p}+T^{*00}_{t}+T^{*00}_{f}$$
and the asterisk denotes a reference frame where the nondiagonal components of the total energy momentum tensor are zero. 
This choice allows to study the influence of hadronic rescatterings to the observables by comparing them with the ones calculated in previous 3-fluid hydrodynamic models.

For the details of fluid to particle conversion the reader is referred to \cite{Ivanov:2005yw}, whereas here we repeat the details important for the construction of the Monte Carlo sampling procedure. 
Both the baryon-rich projectile and target fluids as well as the fireball fluid are being frozen out in small portions, therefore the output of the particlization procedure is a set of droplets (or surface elements). Each droplet is characterized by its proper volume $V^{\rm pr}$, temperature $T$, baryon, $\mu_{\rm B}$, strange chemical potentials $\mu_{\rm S}$, 
and the collective flow velocity $u^\mu$.

The thermodynamic parameters of the droplets correspond to a free hadron resonance gas. 
Therefore, we proceed with sampling the hadrons according to their phase space distributions 
(see Eq.~(33) in \cite{Ivanov:2005yw}), which are expressed in the rest frame of the fluid element (FRF) 
as
\begin{equation}\label{eq-momentum-rf}
p^{*0} \frac{d^3 N_i}{d^3 {\bf p^*}}=\sum\limits_{\alpha} \frac{g_i V_\alpha^{\rm pr}}{(2\pi)^3}
 \frac{p^{*0}}{\exp\left[ (p^{*0} - \mu_{\alpha i}) / T_\alpha \right] \pm 1}
\end{equation}
where the asterisk denotes momentum in the fluid rest frame, where $u^{*\mu}_\alpha=(1,0,0,0)$, 
$\mu_{\alpha i}=B_i\cdot\mu_{\alpha\rm B} + S_i\cdot\mu_{\alpha\rm S}$ is the chemical potential of  hadron $i$ with baryon number $B_i$, strangeness $S_i$, degeneracy factor  $g_i$, 
and the $\alpha$ summation runs over droplets from all (p, t and f) fluids.

The use of temperature and chemical potentials implies a grand canonical ensemble for each surface element.
The sampling is therefore organized as a loop over all droplets, every iteration of which consists of the following steps \cite{Karpenko:2015xea,Karpenko:2013ama}
\begin{itemize}
 \item average multiplicities of all hadron species are calculated according to
 \begin{equation}
  \Delta N_{i,\alpha}=V^{\rm pr}_\alpha n_{i,\rm th}(T, \mu_i),
 \end{equation}
 together with their sum $\Delta N_{\rm tot, \alpha}=\sum_i \Delta N_{i,\alpha}$;
 \item total (integer) number of hadrons from each surface element is sampled according to Poisson distribution with mean $\Delta N_{\rm tot, \alpha}$. If the number is greater than zero, sort of hadron is randomly chosen based on probabilities $\Delta N_{i,\alpha}/\Delta N_{\rm tot, \alpha}$;
 \item hadron's momentum in FRF $p^*$ is sampled according to (\ref{eq-momentum-rf}), which is isotropic in momentum space;
 \item momentum vector is Lorentz boosted to the global frame of the collision.
\end{itemize}
In the present version of the generator, also from the arguments of consistency with preceding hydrodynamic evolution, we do not apply any corrections over the grand canonical procedure to account for effects of charge or energy conservation. 
Therefore, particle multiplicities fluctuate from event to event according to the composition of grand canonical ensembles given by the individual droplets.

\subsection{UrQMD simulation of final state interactions}
\label{UrQMD}

The Ultra-relativistic Quantum Molecular Dynamics (UrQMD) approach \cite{Bass:1993em,Bass:1998ca}
treats
hadrons and resonances up to a mass of $\sim 2.2$ GeV. 
All binary interactions are treated via the excitation and decay of resonances or string 
excitation and decay and elastic scatterings. 
It is crucial for a state-of-the-art event generator to treat the interactions during the late non-equilibrium 
hadronic stage of heavy ion reactions properly. 
At RHIC and LHC notable differences in the proton yields have been observed and the identified particle spectra and flow observables show an effect of the hadronic rescattering 
(for a review of hybrid approaches see \cite{Petersen:2014yqa}). 
At lower beam energies as they are investigated in this work, the hadronic stage of the reaction is of utmost importance. 
In \cite{Auvinen:2013sba} it has been shown, that the excitation function of elliptic and triangular flow can only  be understood within a combined hydrodynamics+transport approach. 
UrQMD constitutes an effective solution of the relativistic Boltzmann equation and therefore provides 
access to the full phase-space distribution of all individual particles at all times. In this work the effect of hadronic rescattering 
in the final state
on the identified particle spectra and the rapidity dependent directed flow is demonstrated in detail. 
%
%
\section{Results}
\label{Applications}
In this section we present a selection of first results from THESEUS for the energy scan ($\sqrt{s_{NN}}=4-11$ GeV) planned at the NICA-MPD collider experiment, which has overlap with the FAIR-SPS-BES/RHIC energy range. 

\begin{figure*}[!th]
\centering
\includegraphics[scale=0.55]{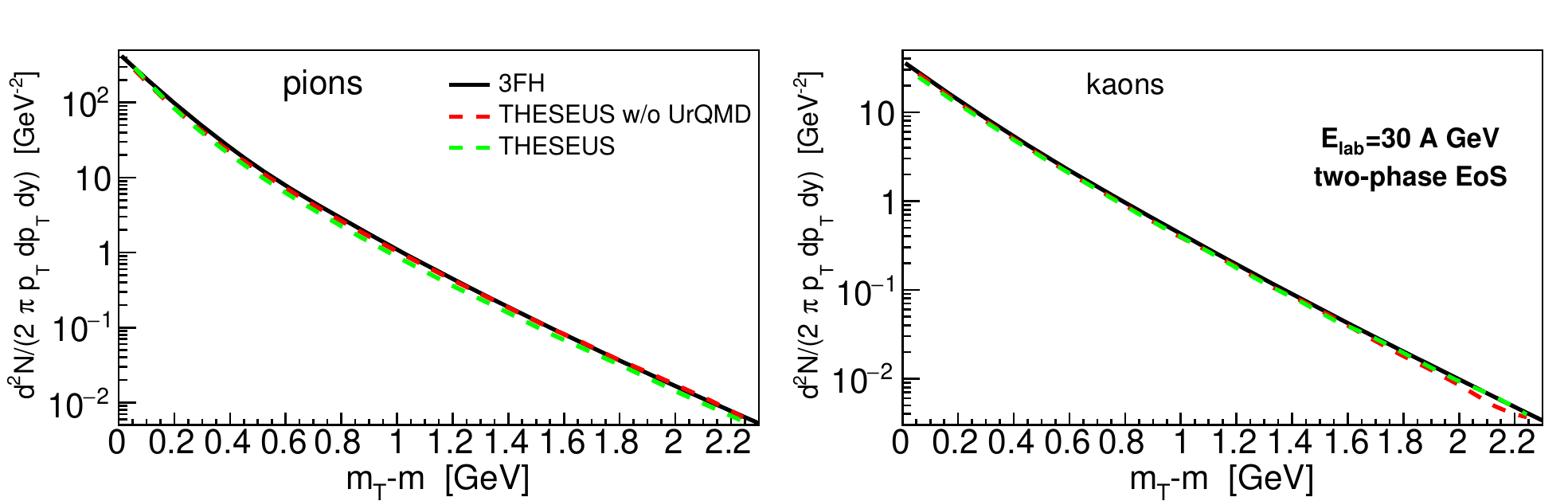}
\caption{Transverse momentum spectrum for pions (left panel) and kaons (right panel) for  central Au+Au collisions ($b=2~$fm) at $E_{\rm lab}=30~$A~GeV for the 2-phase EoS. 
Comparison between results from the 3FH model (black solid lines) and THESEUS without UrQMD (red dashed lines) show excellent agreement.
Comparing these results with the full THESEUS result (green dashed line) shows that the UrQMD hadronic rescattering leads to a slight steepening of the pion $p_T$ spectrum.
\label{fig-pt}}
\end{figure*}

\begin{figure*}[!th]
\centering
\includegraphics[scale=0.6]{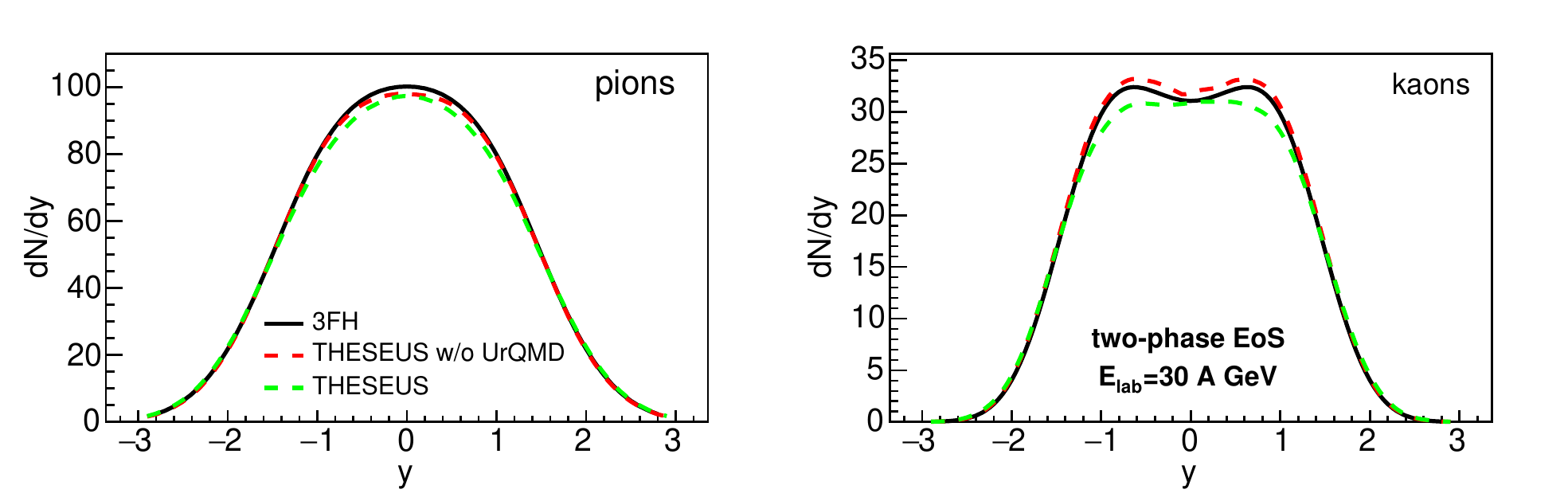}
\caption{Rapidity distribution for pions (left panel) and kaons (right panel) for central Au+Au collisions ($b=2~$fm) at $E_{\rm lab}=30~$A~GeV for the 2-phase EoS. 
Comparison between results from the 3FH model (black solid lines) and THESEUS without UrQMD (red dashed lines) show excellent agreement.
Comparing these results with the full THESEUS result (green dashed line) shows that the UrQMD hadronic rescattering smeares out the double-peak structure in the kaon rapidity spectrum.
\label{fig-dndy}}
\end{figure*}
\subsection{Tests of the particlization routine:
Spectra of pions, kaons and protons}
We start by showing the transverse momentum distributions of pions [$(\pi^+ + \pi^0 + \pi^-)/3$] and kaons [$(K^+ + K^0)/2$] in Fig.~\ref{fig-pt}. 
They are calculated from a sample of 30000 events generated according to the Monte Carlo procedure described above, and are compared in the plot to direct integration according to 
Eq.~(\ref{eq-momentum-rf}) and inclusion of resonance decay contributions performed in the basic 3FH
part of THESEUS (the afterburner is turned off for this comparison). 
The 3FH evolution simulates Au+Au collisions at $E_{\rm lab}=30$~A~GeV with the two-phase EoS. 
We observe excellent agreement up to $p_T=2.2$~GeV, which is limited by the generated event statistics.
In Fig.~\ref{fig-dndy} we show the rapidity distributions for the same setup. 
The rapidity distributions reveal a small difference in kaon yields, and an even smaller one for pions, which is attributed to differences in the large mass sector of the resonance tables and branching ratios between 3FH and THESEUS. Nevertheless, the shapes of rapidity distributions agree beautifully.
In Figs.~~\ref{fig-pt} and \ref{fig-dndy} we show also the effect of the UrQMD hadronic final state interactions
which are included in THESEUS. They lead to a slight steepening of the $p_T$ spectrum for pions and to a
reduction of the double-peak structure in the kaon rapidity spectrum. 
Both are sufficiently gentle effects to not spoil our conclusion.
The tests demonstrate that 
both the procedure of particle sampling
at particlization and the resonance decay kinematics are implemented correctly.

\subsection{Directed flow of protons and pions}
Next we test whether more subtle features of particle distributions are preserved by the particlization procedure and how they are affected by the hadronic cascade. 
First we calculate the directed flow coefficient $v_1$ for pions and protons as a function of rapidity using the reaction plane method
$$v_1(y)=\left< \cos(\phi-\Psi_{\rm RP}) \right>=\left< p_x/\sqrt{p_x^2+p_y^2} \right>,$$
where $\Psi_{\rm RP}=0$ in the model, since the impact parameter is always directed along x-axis.
Although the generator makes it possible to apply different methods of flow analysis over generated events, we use the reaction plane method in order to perform a one-to-one comparison between results from 
THESEUS with and without UrQMD and the corresponding ones from the basic 3FH model.

\begin{figure*}[!h]
\centering
\sidecaption
\includegraphics[scale=0.24]{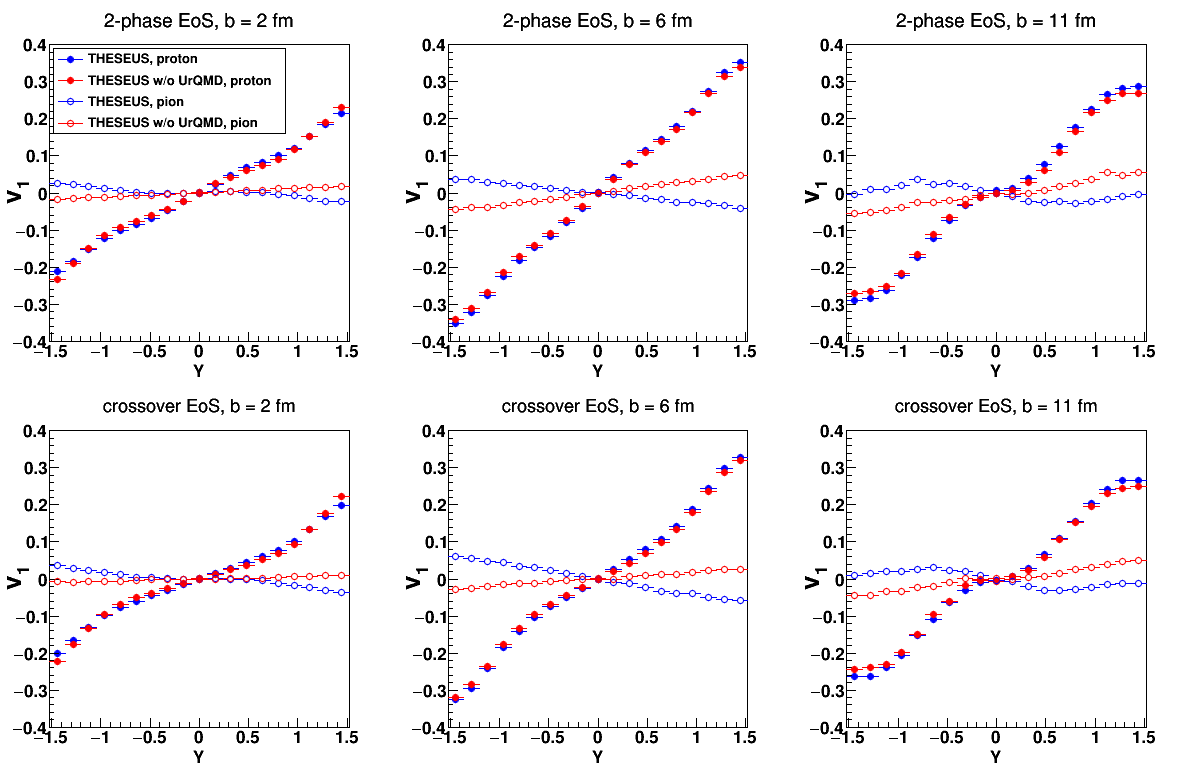}
\caption{
Two upper rows: Directed flow ($v_1$) of protons (full symbols) and pions (open symbols) for central ($b=2$ fm), semicentral ($b=6$ fm) and peripheral ($b=11$ fm) Au+Au collisions at $E_{\rm lab}=8~$A~GeV.  
The upper row is for the 2-phase EoS while the lower row shows results for the crossover EoS.
In each panel we show the direct comparison of THESEUS with (blue symbols) and without (red symbols)
UrQMD afterburner. Remarkable is the effect of turning pion flow to antiflow due to hadronic rescattering 
in the dense baryonic medium.
\label{8AGeV}
}
\end{figure*}
\begin{figure*}[!h]
\centering
\sidecaption
\includegraphics[scale=0.24]{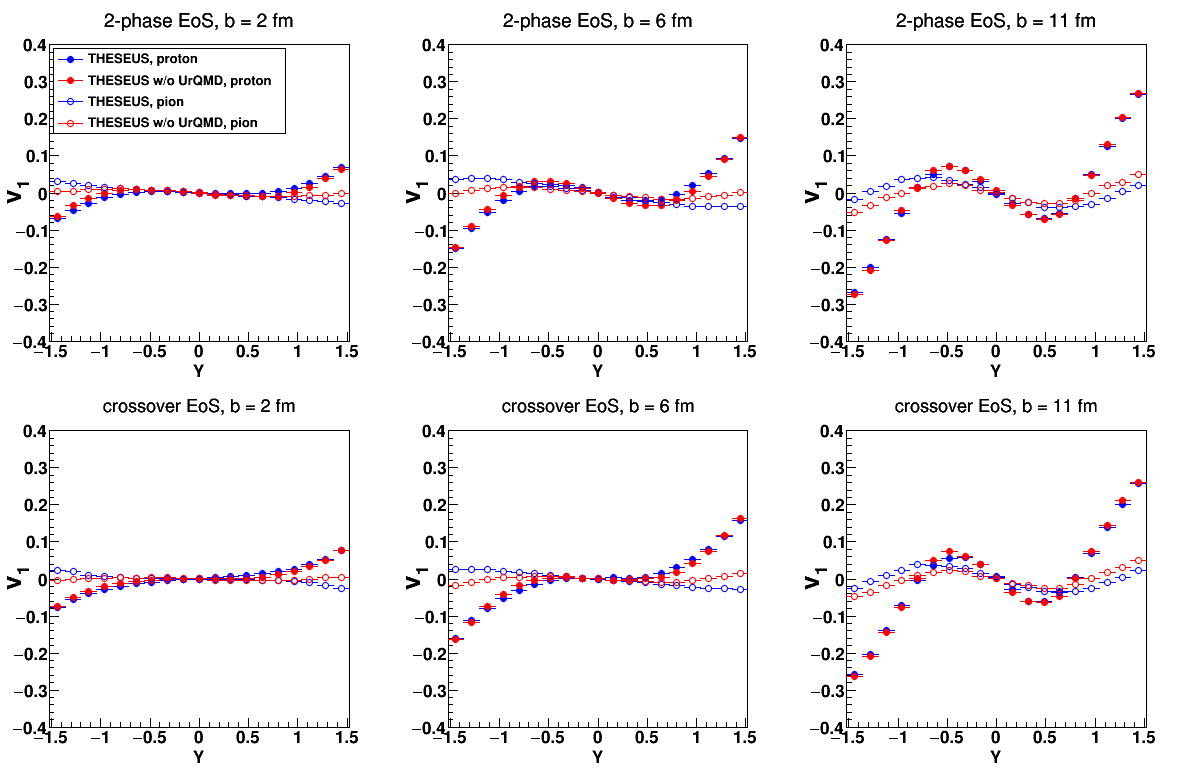}
\caption{
Same as in Fig. \ref{8AGeV} but for $E_{\rm lab}=30~$A~GeV.
\label{30AGeV}
}
\end{figure*}

We present the results of THESEUS with and without UrQMD afterburner for the directed flow 
$v_1$ of protons and pions at $E_{\rm lab}=8~$A~GeV (Fig.~\ref{8AGeV}) and $30~$A~GeV (Fig.~\ref{30AGeV}),  
comparing the case of the 2-phase EoS (first order phase transition, upper panels) with that of the crossover EoS (lower panels) at central (left panels), semicentral (middle panels) and peripheral (right panels) Au+Au collisions. 
Dashed lines show the results from THESEUS without hadronic cascade, where we quantitatively reproduce the results from basic 3FH model.

This figure shows the influence of hadronic final state interactions on the patterns of directed flow of protons and pions in the the broad rapidity range $-1.5 < y < 1.5$ and how it evolves from low energies 
to high energies.
At $E_{\rm lab}=8~$A~GeV in Fig.~\ref{8AGeV} we observe that hadronic rescattering causes the transition from flow to antiflow for pions due to the shadowing by a dense baryonic medium.
The flow of protons is not affected by the hadronic rescattering, which remains so for all energies.
The shadowing effect on the pion directed flow becomes less important at higher energies. 
At $30~$A~GeV hadronic rescattering has no effect on the directed flow of pions in the central rapidity region.

This behaviour can be understood as follows.  
If there is only hydrodynamics, the pions are emitted along the fluid flow, while when there is rescattering they are blocked by the baryonic matter in the projectile and target region, therefore the anti-flow appears. 
This effect was first demonstrated in Ref.~\cite{Bass:1993em}. 
This effect of the pion shadowing is more spectacular in Fig. \ref{8AGeV} where the directed flow 
of protons and pions at $E_{\rm lab}=8~$A~GeV is presented. As seen, the proton $v_1$ is 
practically insensitive to the UrQMD afterburner, while the pion $v_1$ is strongly affected 
by this afterburner. The afterburner even changes the pion $v_1$ flow to an  antiflow. 
The effect of the pion shadowing becomes weaker with the collision energy rise, 
as it is seen at $E_{\rm lab}=30~$A~GeV, because the midrapidity region becomes 
less baryon abundant. 
Though, at 
larger collision energies and peripheral rapidities, 
this shadowing is still noticeable. 
This shadowing results in better agreement with the STAR data on pion $v_1$ 
\cite{Adamczyk:2014ipa}.

There is hardly any difference to be noticed in the pion directed flow patterns between the case of a 2-phase EoS and a crossover EoS.

\begin{figure}[!h]
\centering
\sidecaption
\includegraphics[width=0.3\textwidth]{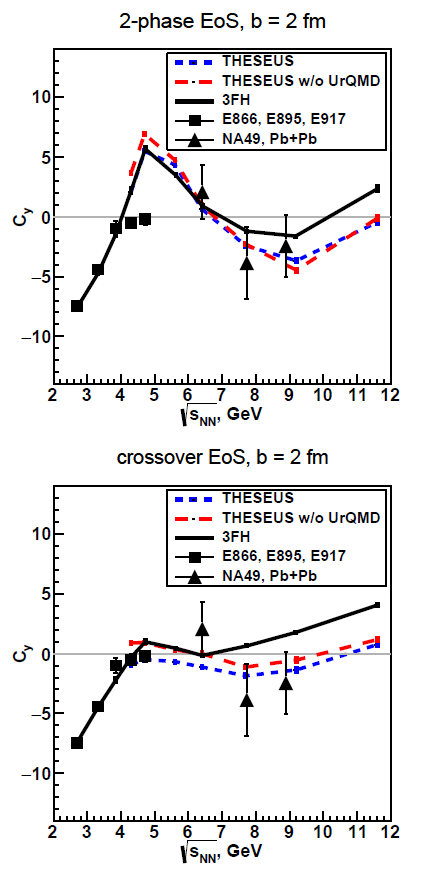}
\caption{ 
Energy scan for the curvature $C_y$ of the net proton rapidity distribution at midrapidity for 
central Au+Au collisions with impact parameter $b=2~$fm. 
We compare the 3FH model result (black solid lines) with THESEUS (blue short-dashed lines) and 
THESEUS without UrQMD (red long-dashed lines).
The results for the two-phase EoS (upper row) are compared to those for the crossover EoS (lower row).  
The "wiggle" as a characteristic feature for the EoS with a first order phase transition is rather robust against  hadronic final state interactions.
Data from AGS experiments are shown by filled squares, data from NA49 by filled triangles.
\label{stopping}
}
\end{figure}

\subsection{Baryon stopping signal for a first-order phase transition}
In Fig.~\ref{stopping} we show the reduced curvature of the net proton rapidity distribution 
$C_y=y_{\rm cm}^2 (d^3 N_{\rm net-p}/dy^3) / (dN_{\rm net-p}/dy)$, where $y_{\rm cm}$ 
is the rapidity of the center of mass of the colliding system in the frame of the target 
\cite{Ivanov:2010cu,Ivanov:2011cb,Ivanov:2015vna}.
Because of a narrower collision energy range chosen here, we observe only the peak-dip part of the
so-called ``peak-dip-peak-dip'' structure reported in \cite{Ivanov:2010cu,Ivanov:2011cb,Ivanov:2015vna}.
The reduced curvature is calculated by fitting the rapidity distribution with a 2$^{\rm nd}$ order polynomial of the form $P_2(y)=ay^2+by+c$ for which then $C_y=y^2_{\rm beam} 2a/c$ results.

Contrary to the basic 3FH model which can calculate $C_y$ with any given precision, in the Monte Carlo procedure the accuracy depends on the event statistics and binning. 
A reliable determination of $C_y$ requires not less than $10^4$ events for central and semi-central collisions and $10^5$ events for peripheral collisions. Larger required statistics for peripheral events is a consequence of the lower average event multiplicity.

The robustness of the baryon stopping signal for a first-order phase transition against experimental cuts in the $p_T$ acceptance has been discussed in \cite{Ivanov:2015vna}.  
It is found that the baryon stopping signal is robust against hadronic rescattering.  

\begin{figure}[!h]
\centering
\includegraphics[width=0.9\textwidth]{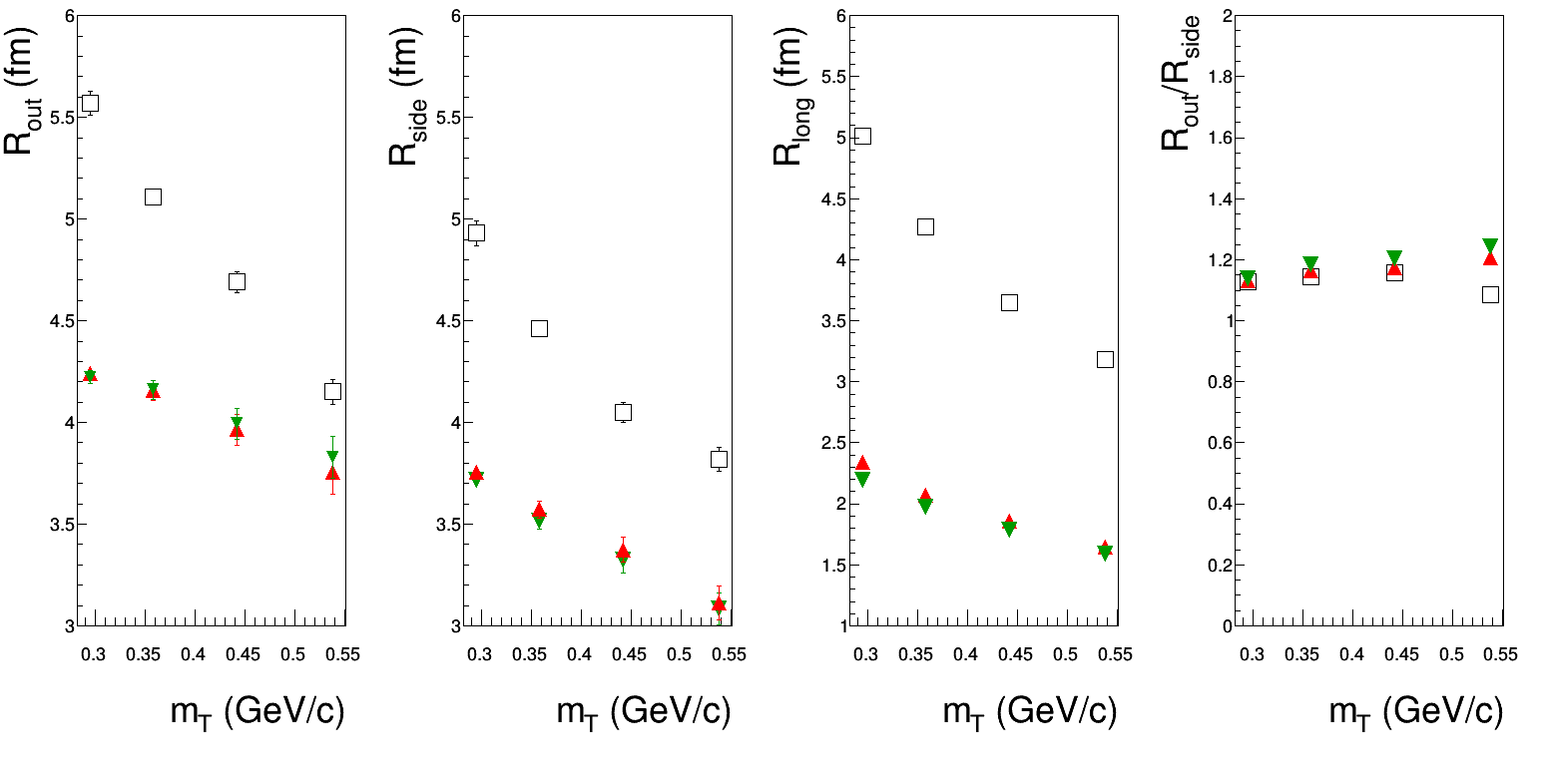}
\caption{ 
Comparison of results of the THESEUS without UrQMD afterburner on femtoscopy radii 
with those measured
by the STAR collaboration at $\sqrt{s_{NN}} =$ 7.7 GeV (opened squares). Triangles correspond to different types of EoS: the two-phase EoS (reversed triangles) and the crossover EoS (triangles).
\label{femto-radii-wo-UrQMD}
}
\end{figure}

\begin{figure}[!h]
\centering
\includegraphics[width=0.9\textwidth]{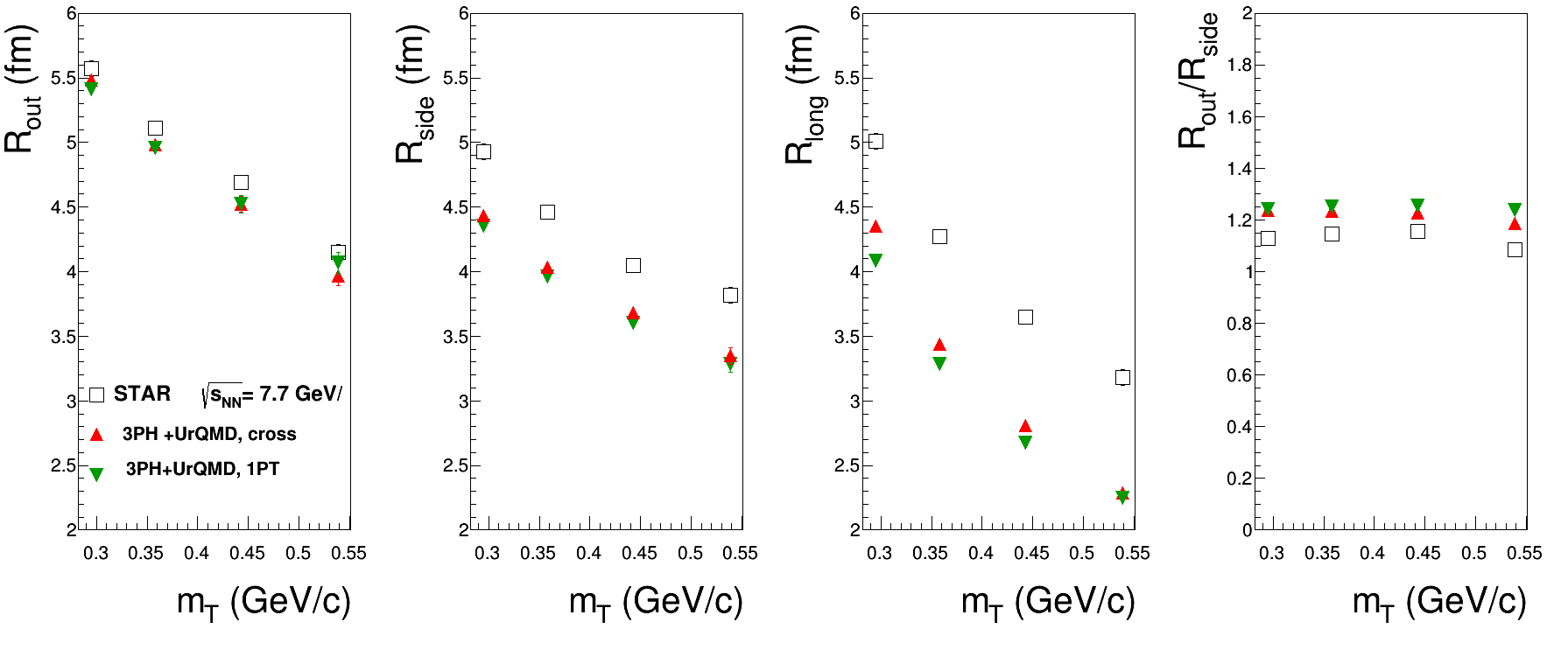}
\caption{ 
The same as in Fig. \ref{femto-radii-wo-UrQMD} but with the UrQMD afterburner. 
\label{femto-radii+UrQMD}
}
\end{figure}

\subsection{Preliminary results on femtoscopy}
Correlation femtoscopy allows one to measure the space-time characteristics of particle production
in relativistic heavy-ion collisions due to the effects of quantum statistics and final state
interactions. The femtoscopy was intensively
studied at AGS and SPS accelerators and is being studied now at the BES/RHIC 
in the context of exploration of the QCD
phase diagram. In this contribution we present preliminary results on femtoscopy observables 
 for central Au-Au collisions at
$\sqrt{s_{NN}} =$ 7.7 GeV calculated by means of the event generator THESEUS. 

The femtoscopic analysis within THESEUS was performed very similarly to that described in 
Ref. \cite{Batyuk:2017smw}. 
A two-pion correlation function is fitted by the conventional Gaussian form
   \begin{eqnarray}
   \label{2p-corr}
C({\bf q}) = N [
1 + \lambda \; exp(-R^2_{out}\,q^2_{out} - R^2_{side}\,q^2_{side} - R^2_{long}\,q^2_{long})]
   \end{eqnarray}
where 
${\bf q} = {\bf p}_1 - {\bf p}_2$ with
${\bf p}_1$ and ${\bf p}_2$ being three-momenta of two considered particles, 
$N$ is the normalization factor and $\lambda$ is the correlation strength parameter, 
which can differ from unity
due to the contribution of long-lived emitters and a non-Gaussian shape of the correlation function; 
$R_{out}$, $R_{side}$ and $R_{long}$ are the Gaussian femtoscopy radii in the in the longitudinally co-moving 
system (LCMS), where the longitudinal pair momentum vanishes.

Preliminary results on $R_{out}$, $R_{side}$ and $R_{long}$ as functions of 
the transverse mass of the particles, $m_T = \sqrt{({\bf p}_1+{\bf p}_2)^2/4+m_\pi}$,
for central Au-Au collisions at
$\sqrt{s_{NN}} =$ 7.7 GeV simulated by means of THESEUS without the UrQMD afterburner
are presented in Fig.  \ref{femto-radii-wo-UrQMD}. The corresponding experimental data 
\cite{Adamczyk:2014mxp} are also displayed in Fig.  \ref{femto-radii-wo-UrQMD}. 
As seen, the femtoscopy radii are strongly underestimated without the afterburner stage. 

Figure \ref{femto-radii+UrQMD} demonstrates these radii after the afterburner stage. 
Now these radii are already in a reasonable agreement with the STAR data. 
The  $R_{out}/R_{side}$ ratio is even in a very good agreement with the data. 
An unexpected result is that the femtoscopy radii turn out to be very close 
within the first-order-transition and crossover scenarios. 
The analysis of the femtoscopy observables is still in progress.

\section{Conclusions}
\label{Conclusions}
We have assembled the new event generator THESEUS that is based on a three-fluid hydrodynamics description of the early and dense stages of the collision, followed by a particlization 
as input to the 
UrQMD "afterburner" accounting for hadronic final state interactions. 
The new simulation program has the unique feature that it can describe a hadron-to-quark matter transition 
which proceeds in the baryon stopping regime that is not accessible to previous simulation programs that are designed for higher energies.

We presented first results from THESEUS for the FAIR-NICA-SPS-BES/RHIC energy scan addressing the directed flow of protons and pions as well as the rapidity distribution of protons, pions and kaons for two model EoS, 
one with a first order phase transition, the other with a crossover type softening at high densities. 
Preliminary results on the femtoscopy are also discussed. 
Another important application of THESEUS, not discussed in the present contribution, is  
prediction of the directed flow of deuterons in semicentral Au+Au collisions in the NICA energy range \cite{Bastian:2016xna}.  In Ref. \cite{Bastian:2016xna} it is argued that 
light clusters are  unique  rare probes of in-medium characteristics such as phase space occupation and early flow.

We have found that the hadronic cascade which is switched on after the particlization has little effect 
on the proton flow observables and on  
the predicted 
baryon stopping signal for a first-order phase transition in heavy-ion collisions at NICA/FAIR energies. 
However, for pions in non-central collisions at lower energies the hadronic cascade leads to a qualitative change of the emission pattern (from flow to antiflow).
The femtoscopy observables are strongly affected by the afterburner stage.
The preliminary analysis of the femtoscopy radii manifested their reasonable 
agreement with the STAR data.

%
The work of P.B., D.B., Yu.B.I., S.M. and O.R. 
was supported by the Russian Science Foundation, Grant No. 17-12-01427.
M.B. was partially supported by COST Action CA15213 (THOR). 
Iu.K. acknowledges partial support by the University of Florence
grant {\it Fisica dei plasmi relativistici: teoria e applicazioni moderne}. 
H.P. acknowledges funding of a Helmholtz Young Investigator Group VH-NG-822 from
the Helmholtz Association and GSI.
M.N. acknowledges fruitful discussions within the framework of the Beam Energy Scan
Theory (BEST) Topical Collaboration. 
%


\begin{thebibliography}{999}
%
\bibitem{Stephans:2006tg}
G.~S.~F.~Stephans,
  J.\ Phys.\ G {\bf 32}, S447 (2006).
%
\bibitem{SPS-scan}
P. Seyboth [NA49 Collaboration], Addendum-1 to the NA49 Proposal, CERNSPSC-
97-26;
M.~Gazdzicki, nucl-th/9701050; 
M.~Gazdzicki {\it et al.} [NA61/SHINE Collaboration], PoS C POD2006, 016 (2006).
%
\bibitem{FAIR}  
  B.~Friman, (ed.) {\it et al.}, 
  Lect.\ Notes Phys.\  {\bf 814}, 1 (2011).
%
\bibitem{NICA} 
  A.~N.~Sissakian, A.~S.~Sorin and V.~D.~Toneev,
  Conf.\ Proc.\ C {\bf 060726}, 421 (2006).
%
%
\bibitem{Ivanov:2005yw} 
  Yu.~B.~Ivanov, V.~N.~Russkikh and V.~D.~Toneev,
  Phys.\ Rev.\ C {\bf 73}, 044904 (2006).
%
\bibitem{Ivanov:2013wha} 
  Yu.~B.~Ivanov,
  Phys.\ Rev.\ C {\bf 87}, no. 6, 064904 (2013).
%
\bibitem{Ivanov:2012bh} 
  Yu.~B.~Ivanov,
  Phys.\ Lett.\ B {\bf 721}, 123 (2013).
%
\bibitem{Ivanov:2013yqa} 
  Yu.~B.~Ivanov,
  Phys.\ Rev.\ C {\bf 87}, no. 6, 064905 (2013).
%
\bibitem{Ivanov:2013yla} 
  Yu.~B.~Ivanov,
  Phys.\ Rev.\ C {\bf 89}, no. 2, 024903 (2014).
%
\bibitem{Konchakovski:2014gda} 
  V.~P.~Konchakovski {\it et al.}, 
  Phys.\ Rev.\ C {\bf 90}, no. 1, 014903 (2014).
%
\bibitem{Ivanov:2014zqa} 
  Yu.~B.~Ivanov and A.~A.~Soldatov,
  Phys.\ Rev.\ C {\bf 91}, no. 2, 024914 (2015).
%
\bibitem{Ivanov:2014ioa} 
  Yu.~B.~Ivanov and A.~A.~Soldatov,
  Phys.\ Rev.\ C {\bf 91}, no. 2, 024915 (2015).
%
\bibitem{Ivanov:2016sqy} 
  Yu.~B.~Ivanov and A.~A.~Soldatov,
  Eur.\ Phys.\ J.\ A {\bf 52}, no. 1, 10 (2016).
%
\bibitem{Batyuk:2016qmb} 
  P.~Batyuk {\it et al.},
  Phys.\ Rev.\ C {\bf 94}, 044917 (2016).
%
\bibitem{Bass:1993em} 
  S.~A.~Bass, R.~Mattiello, H.~St\"ocker, W.~Greiner and C.~Hartnack,
  Phys.\ Lett.\ B {\bf 302}, 381 (1993).
%
\bibitem{Bass:1998ca} 
  S.~A.~Bass {\it et al.},
  Prog.\ Part.\ Nucl.\ Phys.\  {\bf 41}, 255 (1998).
\bibitem{Hung:1997du}
  C.~M.~Hung and E.~V.~Shuryak,
  Phys.\ Rev.\ C {\bf 57}, 1891 (1998).
%
\bibitem{Ivanov:2015vna} 
  Yu.~B.~Ivanov and D.~Blaschke,
  Phys.\ Rev.\ C {\bf 92}, 024916 (2015).
%
%
\bibitem{Mishustin:1988mj} 
  I.~N.~Mishustin, V.~N.~Russkikh and L.~M.~Satarov,
  Sov.\ J.\ Nucl.\ Phys.\  {\bf 48}, 454 (1988).
%
\bibitem{Mishustin:1991sp} 
  I.~N.~Mishustin, V.~N.~Russkikh and L.~M.~Satarov,
  Sov.\ J.\ Nucl.\ Phys.\  {\bf 54}, 260 (1991).
%
\bibitem{Katscher:1993xs} 
  U.~Katscher {\it et al.}, 
  Z.\ Phys.\ A {\bf 346}, 209 (1993).
%
\bibitem{Brachmann:1997bq} 
  J.~Brachmann {\it et al.}, 
  Nucl.\ Phys.\ A {\bf 619}, 391 (1997).
%
%
\bibitem{gasEOS}
  V.~M.~Galitsky and I.~N.~Mishustin,
  Yad.\ Fiz.\  {\bf 29}, 363 (1979).
%
\bibitem{Toneev06}
  A.~S.~Khvorostukin, V.~V.~Skokov, V.~D.~Toneev and K.~Redlich,
  Eur.\ Phys.\ J.\ C {\bf 48}, 531 (2006).
%
%
\bibitem{71}
       V. N. Russkikh, Yu. B. Ivanov, 
       Phys. Rev. C {\bf 76}, 054907 (2007). 
%
\bibitem{74}
       Yu. B. Ivanov, V. N. Russkikh,
       Phys. Atom. Nucl. {\bf 72}, 1238 (2009).
\bibitem{Huovinen:2012is} 
  P.~Huovinen and H.~Petersen,
  Eur.\ Phys.\ J.\ A {\bf 48}, 171 (2012).

\bibitem{Karpenko:2015xea} 
  I.~A.~Karpenko {\it et al.}, 
  Phys.\ Rev.\ C {\bf 91}, no. 6, 064901 (2015).


\bibitem{Karpenko:2013ama} 
  I.~A.~Karpenko {\it et al.}, 
  J.\ Phys.\ Conf.\ Ser.\  {\bf 503}, 012040 (2014).
%
\bibitem{Petersen:2014yqa} 
  H.~Petersen,
  J.\ Phys.\ G {\bf 41}, no. 12, 124005 (2014).

\bibitem{Auvinen:2013sba} 
  J.~Auvinen and H.~Petersen,
  Phys.\ Rev.\ C {\bf 88}, no. 6, 064908 (2013).
%
\bibitem{Liu:2000am} 
  H.~Liu {\it et al.} [E895 Collaboration],
  Phys.\ Rev.\ Lett.\  {\bf 84}, 5488 (2000).
%
\bibitem{Adamczyk:2014ipa} 
  L.~Adamczyk {\it et al.} [STAR Collaboration],
  Phys.\ Rev.\ Lett.\  {\bf 112}, no. 16, 162301 (2014).
%
\bibitem{Alt:2003ab} 
  C.~Alt {\it et al.} [NA49 Collaboration],
  Phys.\ Rev.\ C {\bf 68}, 034903 (2003).
%
\bibitem{Ivanov:2010cu}
  Yu.~B.~Ivanov,
  Phys.\ Lett.\  B {\bf 690}, 358 (2010).
%
\bibitem{Ivanov:2011cb} 
  Yu.~B.~Ivanov,
Phys. At. Nucl. {\bf 75} 621 (2012).  
%
\bibitem{Batyuk:2017smw} 
  P.~Batyuk {\it et al.}, 
  Phys.\ Rev.\ C {\bf 96}, no. 2, 024911 (2017).
%
\bibitem{Adamczyk:2014mxp} 
  L.~Adamczyk {\it et al.} [STAR Collaboration],
  Phys.\ Rev.\ C {\bf 92}, no. 1, 014904 (2015).
%
\bibitem{Bastian:2016xna} 
  N.-U.~Bastian {\it et al.},
  Eur.\ Phys.\ J.\ A {\bf 52}, no. 8, 244 (2016). 
\end{thebibliography}
\end{document}